# Power Transmittance of a Laterally Shifted Gaussian Beam through a Circular Aperture

**Tariq Shamim Khwaja[1] and Syed Azer Reza[1]**

1. Department of Electrical Engineering, Lahore University of Management Sciences, DHA, Lahore 54792, Pakistan

**Abstract**

Gaussian beams are often used in optical systems. The fundamental Gaussian $TEM_{00}$ mode is the most common of the Gaussian modes present in various optical devices, systems and equipment. Within an optical system, it is common that this Gaussian $TEM_{00}$ beam passes through a circular aperture of a finite diameter. Such circular apertures include irises, spatial filters, circular Photo-Detectors (PDs) and optical mounts with circular rims. The magnitude of optical power passing through a finite-sized circular aperture is well-documented for cases where the Gaussian beam passes through the center of the clear circular aperture, and is chopped off symmetrically in all radial directions on a given plane. More often than not, a non-axial incident Gaussian Beam is not blocked in a radially uniform manner by a circular aperture. Such situations arise due to a lateral displacement of the beam from tilted glass blocks, manufacturing errors and imperfect surface flatness or parallelness of surfaces. The fraction of optical power of a laterally-shifted Gaussian Beam passing through a circular aperture is calculated in this paper through conventional integration techniques.

## 1. Introduction

In most laser-based optical systems, beams a Laser Source (LS) pass through some optical components with circular apertures such as spherical lenses, circular mirrors, irises and circular spatial filters, circular gratings, optical wavelength filters, attenuators and many more. Moreover, Photo-Detectors (PDs), used for detection of the incident optical power, commonly have a circular active area. Typically the various optical elements inside an optical system are aligned so that the optical axis of the system is coincident with the optical axis of individual elements such as lenses and curved mirrors which allows the laser beam to propagate through the center of these elements. In this case, all elements in the optical system are rotationally symmetric about the optical axis of the system.

Most laser sources such as He-Ne lasers emit the fundamental Gaussian Mode $TEM_{00}$. When a Gaussian Beam propagates through optical elements with circular apertures, a fraction of the total beam power is blocked due to a finite clear aperture radius of each element. In the case of a PD with a circular active area, only the fraction of optical power which is incident on the PD active area contributes to the generated photo-current while the remaining optical power is not recorded.

The transmitted power of a Gaussian beam passing through a circular aperture is well-documented for the case when the center of the beam irradiance profile coincides with the center of the aperture [1]. In such a case, the two-dimensional irradiance profile function of the Gaussian Beam is integrated over the dimensions of the clear aperture of the circular optical element and multiplied, depending on beam polarization, by the Fresnel Power Transmission Coefficient [2] to determine the power transmitted. As the centers of the beam profile and aperture coincide, the integration is rather simple due to a radially symmetric integral.

On the other hand, the calculation of transmitted optical power of a Gaussian Beam becomes tedious when the beam is laterally displaced with respect to the center of clear aperture of an optical element. A simpler case of power transmission or detection of a laterally-shifted beam through a square aperture has been presented in [3]. In the case of a circular aperture, the integration of the two-dimensional Gaussian irradiance function over a circular clear aperture is not radially symmetric which renders it a non-trivial integral to solve. The effect on transmitted optical power due to a lateral beam shift is negligible when the clear aperture diameter is much larger than the Full Width Half of Maximum (FWHM) beam diameter as well as the magnitude of lateral shift. But transmitted power can be severely affected by small lateral shifts if the aperture diameter is comparable to the FWHM beam diameter. This is often the case when beams are passing through small irises or spatial filters or if beams are incident on PDs with small active area diameters.

Lateral displacement of a Gaussian beam is very common within optical setups where it can be a result of lateral or tilt misalignments or defects in optical components. Also such a lateral beam displacement can be the result of plates which are often designed as wedge-shaped to mitigate back-reflection effects. It is also known that tunable focus lenses often exhibit a non-parallelism of two surfaces that hold liquid between them, resulting in an angular deviation of the beam which passes through them [4]. This angular deviation results in a lateral beam shift at any plane after the lens.

In this paper, we calculate the percentage transmitted optical power of a beam which is laterally shifted with respect to the center of a circular aperture. In the case of photo-detection with circular detectors, this will signify the percentage of incident optical power which is recorded by a PD.

## 2. Transmittance of a Non-Centered Gaussian Beam through a Circular Aperture

Fig.1 shows a Gaussian Beam which is incident such that its peak of irradiance profile is coincident with the center of a circular aperture. We also show in Fig.1, a laterally displaced Gaussian Beam with respect to the center of a circular clear aperture. In this section, we aim to determine the transmitted optical power in the latter case.

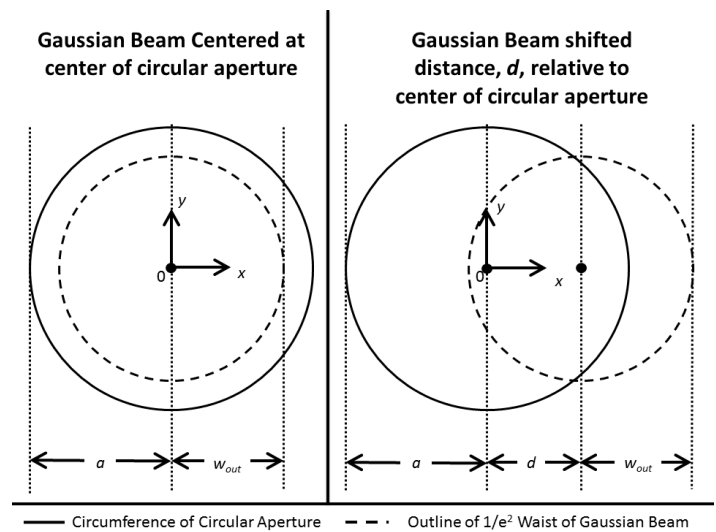

Fig.1 Gaussian Beam incidence on-center and off-center of a circular aperture.

The irradiance profile, $I(x', y', z)$, in Cartesian Coordinates $(x', y', z)$, of a Gaussian beam with minimum beam waist $w_0$ at $z = 0$, is given by:

$$I(x', y', z) = I_{Peak} \left(\frac{w_0}{w(z)}\right)^2 \exp\left(-\frac{2((x')^2 + (y')^2)}{w^2(z)}\right)$$

(1)

If the plane of the circular aperture is located at $z = z'$, then $w(z') = w_{out}$ is the $1/e^2$ beam waist radius and $I_{Peak}$ is the peak irradiance at the center of the Gaussian Beam at this location. From Eq.(1), the irradiance profile of the Gaussian beam at $z = z'$ is given by:

$$I(x, y, z) = I_{Peak} \left(\frac{w_0}{w_{out}}\right)^2 \exp\left(-\frac{2((x')^2 + (y')^2)}{w_{out}^2}\right)$$

(2)

As shown in Fig.1, we choose the coordinates $(x,y)$ according to the circular aperture such that the point $(0,0)$ signifies the center of the aperture. The lateral displacement of the center of the Gaussian Beam is $d$. As the choice of primary coordinate axes is arbitrary, the direction of beam displacement from the center location is taken to be the x-axis. The irradiance profile in Eq.(2) is then expressed in $(x,y)$ coordinates as:

$$I(x, y, z) = I_{Peak} \left(\frac{w_0}{w_{out}}\right)^2 \exp\left(-\frac{2((x-d)^2 + y^2)}{w_{out}^2}\right) = I_0 \left(\frac{w_0}{w_{out}}\right)^2 \exp\left(-\frac{2(x^2 + y^2 + d^2 - 2xd)}{w_{out}^2}\right)$$

(3)

Which can be simplified to

$$I(x, y, z) = I_{Peak} \left(\frac{w_0}{w_{out}}\right)^2 \exp\left(-\frac{2x^2 + 2y^2}{w_{out}^2}\right) \exp\left(-\frac{2d^2}{w_{out}^2}\right) \exp\left(\frac{4xd}{w_{out}^2}\right)$$

(4)

When expressed in cylindrical coordinates (ρ, θ, z), Eq.(4) may be written as:

$$I(x, y, z) = I_{Peak} \left(\frac{w_0}{w_{out}}\right)^2 \exp\left(-\frac{2\rho^2}{w_{out}^2}\right) \exp\left(-\frac{2d^2}{w_{out}^2}\right) \exp\left(\frac{4\rho d\cos\theta}{w_{out}^2}\right)$$

(5)

To find the optical power $P_T$ transmitted through a circular aperture, we integrate the irradiance of the displaced beam over the entire aperture of radius $a$:

$$P_T = \int_0^a \int_0^{2\pi} I_{Peak} \left(\frac{w_0}{w_{out}}\right)^2 \exp\left(-\frac{2\rho^2}{w_{out}^2}\right) \exp\left(-\frac{2d^2}{w_{out}^2}\right) \exp\left(\frac{4\rho d \cos\theta}{w_{out}^2}\right) d\theta\, \rho d\rho$$

(6)

$$\Rightarrow P_T = I_{Peak} \left(\frac{w_0}{w_{out}}\right)^2 \exp\left(-\frac{2d^2}{w_{out}^2}\right) \int_0^a \rho \exp\left(-\frac{2\rho^2}{w_{out}^2}\right) \int_0^{2\pi} \exp\left(\frac{4\rho d \cos\theta}{w_{out}^2}\right) d\theta\, d\rho$$

(7)

$$\Rightarrow P_T = I_{Peak} \left(\frac{w_0}{w_{out}}\right)^2 \exp\left(-\frac{2d^2}{w_{out}^2}\right) 2\pi \int_0^a \rho \exp\left(-\frac{2\rho^2}{w_{out}^2}\right) I_0\left(\frac{4\rho d}{w_{out}^2}\right) d\rho$$

(8)

Where $I_0(x)$ is the Modified Bessel Function of the first kind and order 0 which can be expanded to obtain an expression for $P_T$ as:

$$P_T = I_{Peak} \left(\frac{w_0}{w_{out}}\right)^2 \exp\left(-\frac{2d^2}{w_{out}^2}\right) 2\pi \int_0^a \rho \exp\left(-\frac{2\rho^2}{w_{out}^2}\right) \sum_{k=0}^{\infty} \frac{\left(\frac{1}{4}\left(\frac{4\rho d}{w_{out}^2}\right)^2\right)^k}{(k!)^2} d\rho$$

(9)

$$\Rightarrow P_T = I_{Peak} \left(\frac{w_0}{w_{out}}\right)^2 \exp\left(-\frac{2d^2}{w_{out}^2}\right) 2\pi \int_0^a \rho \exp\left(-\frac{2\rho^2}{w_{out}^2}\right) \sum_{k=0}^{\infty} \frac{4^k d^{2k}}{w_{out}^{4k}(k!)^2} \rho^{2k} d\rho$$

(10)

$$\Rightarrow P_T = I_{Peak} \left(\frac{w_0}{w_{out}}\right)^2 \exp\left(-\frac{2d^2}{w_{out}^2}\right) 2\pi \sum_{k=0}^{\infty} \frac{4^k d^{2k}}{w_{out}^{4k}(k!)^2} \int_0^a \rho^{2k+1} \exp\left(-\frac{2\rho^2}{w_{out}^2}\right) d\rho$$

(11)

Solving the remaining integral results in:

$$P_T = I_{Peak} \left(\frac{w_0}{w_{out}}\right)^2 \exp\left(-\frac{2d^2}{w_{out}^2}\right) 2\pi \sum_{k=0}^{\infty} \frac{4^k d^{2k}}{w_{out}^{4k}(k!)^2}$$
$$\times \left[a^{2k} 2^{-k-2} w_{out}^2 \left(\frac{a^2}{w_{out}^2}\right)^{-k} \left(\Gamma(k+1) - \Gamma\left(k+1, \frac{2a^2}{w_{out}^2}\right)\right)\right]$$

(12)

where $\Gamma(s)$ is the standard Gamma Function and $\Gamma(s,f)$ is the upper incomplete Gamma Function. Simplifying and rearranging terms in Eq.(12) leads to:

$$P_T = 2\pi I_{Peak} \left(\frac{w_0}{w_{out}}\right)^2 \exp\left(-\frac{2d^2}{w_{out}^2}\right)$$
$$\times \sum_{k=0}^{\infty} \frac{4^k d^{2k}}{w_{out}^{4k}(k!)^2} \left[a^{2k} 2^{-k-2} w_{out}^2 \frac{a^{-2k}}{w_{out}^{-2k}} \left(\Gamma(k+1) - \Gamma\left(k+1, \frac{2a^2}{w_{out}^2}\right)\right)\right]$$

(13)

$$P_T = 2\pi I_{Peak} \left(\frac{w_0}{w_{out}}\right)^2 \exp\left(-\frac{2d^2}{w_{out}^2}\right)$$
$$\times \sum_{k=0}^{\infty} \frac{4^k d^{2k}}{w_{out}^{4k}(k!)^2} \left[2^{-k-2} w_{out}^{2k+2} \left(\Gamma(k+1) - \Gamma\left(k+1, \frac{2a^2}{w_{out}^2}\right)\right)\right]$$

(14)

We also know from [5] the relationship between the lower and upper incomplete Gamma Functions and the Gamma Function

$$\gamma\left(k+1, \frac{2a^2}{w_{out}^2}\right) = \Gamma(k+1) - \Gamma\left(k+1, \frac{2a^2}{w_{out}^2}\right)$$

(15)

Where $\gamma(a,x)$ is the lower incomplete Gamma Function. A simplified expression for the optical power of a Gaussian Beam transmitted through a circular aperture is:

$$P_{opt} = \frac{\pi w_0^2 I_{Peak}}{2} \exp\left(-\frac{2d^2}{w_{out}^2}\right) \sum_{k=0}^{\infty} \frac{2^k d^{2k}}{w_{out}^{2k}(k!)^2} \left(\gamma\left(k+1, \frac{2a^2}{w_{out}^2}\right)\right)$$

(16)

For the case when the lateral displacement of the Gaussian beam is zero, the beam passes through the center of the aperture, i.e. $d = 0$. We know that $d^{2k} \neq 0$ only for $k = 0$. For this case the summation in Eq.(15) simplifies and we obtain:

$$P_{opt} = 2\pi I_{Peak} \left(\frac{w_0}{w_{out}}\right)^2 \frac{2^{0-2}}{w_{out}^{0-2}(0!)^2} \left(\gamma\left(1, \frac{2a^2}{w_{out}^2}\right)\right)$$

(17)

As we know that

$$\gamma\left(1, \frac{2a^2}{w_{out}^2}\right) = 1 - \exp\left(-\frac{2a^2}{w_{out}^2}\right)$$

(18)

Substitution of Eq.(18) in Eq.(19) leads to the well-established expression for the transmitted optical power for a non-displaced Gaussian Beam.

$$P_{opt} = \frac{\pi w_0^2 I_{Peak}}{2}\left(1 - \exp\left(-\frac{2a^2}{w_{out}^2}\right)\right)$$

(19)

The use of Eq.(16) to obtain this known expression for transmitted power through a circular aperture for a non-displaced beam demonstrates the validity of Eq.(16). The total power $P_{total}$ of the Gaussian beam is:

$$P_{total} = \int_0^\infty \int_0^{2\pi} I_{Peak} \left(\frac{w_0}{w_{out}}\right)^2 \exp\left(-\frac{2\rho^2}{w_{out}^2}\right) \exp\left(-\frac{2d^2}{w_{out}^2}\right) \exp\left(\frac{4\rho d \cos\theta}{w_{out}^2}\right) d\theta\, \rho d\rho = \frac{\pi w_0^2 I_{Peak}}{2}$$

(20)

Substituting Eq.(20) in Eq.(16), we obtain an expression for power transmittance $T$ of a laterally displaced Gaussian Beam through a circular aperture as:

$$T = \frac{P_{Transmitted}}{P_{total}} = \exp\left(-\frac{2d^2}{w_{out}^2}\right) \sum_{k=0}^\infty \frac{2^k d^{2k}}{w_{out}^{2k}(k!)^2} \gamma\left(k+1, \frac{2a^2}{w_{out}^2}\right)$$

(21)

## 3. Conclusion

In this paper, the authors have presented a mathematical solution to a fundamental and common problem in optical systems. Most optical systems rely on Gaussian $TEM_{00}$ mode propagation and it is commonly assumed that power transmittance through a circular aperture is either very close to 100% for apertures radii which are much larger than the Gaussian Beam waist. For smaller circular apertures, the centers of the beam irradiance distribution and the circular apertures are most assumed to coincide to simplify calculation of transmitted optical power through the aperture. In this paper we have presented an accurate expression for power transmission of a laterally shifted Gaussian Beam through a circular aperture. The expression explains a variation in power transmission for any lateral shifted Gaussian Beam and it simplifies to a known relationship when the displacement is set to zero.

## References

1. A. E. Siegman, "Lasers," pp. 666, University Science, 1986.

2. E. Hecht, "Optics," 4th edition, Addison Wesley, 2002.

3. Arshad, Muhammad Assad, Syed Azer Reza, and Ahsan Muhammad. "Data transfer through beam steering using agile lensing." In SPIE Photonics Europe, pp. 988929-988929. International Society for Optics and Photonics, 2016.

4. Parrot SA Confidential, "Arctic 316 Arctic 316-AR850," 2016.

5. Davis, Philip J. "Gamma function and related functions." Handbook of Mathematical Functions (M (1965).